\documentclass{article}

\usepackage{soul}
\usepackage{booktabs}
\usepackage{multirow}
\usepackage{wrapfig} 
\usepackage{multirow}
\usepackage{etoc}
\usepackage{listings}
\usepackage{fancybox}
\usepackage{enumitem}
\usepackage{graphicx}
\usepackage{color}
\usepackage{array}
\usepackage{amsmath}
\usepackage{xspace}
\usepackage{url}
\usepackage{tikz}
\usepackage{caption}
\usepackage{pgfplots}
\usepackage{balance}
\usepackage{subcaption}
\pgfplotsset{compat=1.16}
\usepackage{pgf-pie}
\usetikzlibrary{pgfplots.statistics,calc}
\usepackage{algorithm}
\usepackage{algorithmic}
\usepackage{adjustbox}
\usepackage{makecell}

\usepackage{tikz}

\usepackage{tcolorbox}
\makeatletter
\newcommand{\mybox}[1]{%
	\setbox0=\hbox{#1}%
	\setlength{\@tempdima}{\dimexpr\wd0+13pt}%
	\begin{tcolorbox}[boxrule=0.5pt, colback=white, arc=4pt,
		left=6pt,right=6pt,top=6pt,bottom=6pt,boxsep=0pt]
		#1
	\end{tcolorbox}
}

\definecolor{songcolor}{RGB}{191,191,191}

\newcommand{\testgen}{\textsc{NeuroTestGen}}
\newcommand{\tool}{\textsc{NeuroTestGen}}

\definecolor{codegreen}{rgb}{0,0.6,0}
\definecolor{codegray}{rgb}{0.5,0.5,0.5}
\definecolor{codepurple}{rgb}{0.58,0,0.82}
\definecolor{backcolour}{rgb}{0.95,0.95,0.92}

\lstdefinestyle{mystyle}{
  language=Java,
  aboveskip=3mm,
  showstringspaces=false,
  columns=flexible,
  numbers=none,
  backgroundcolor=\color{backcolour},
  commentstyle=\color{codegreen},
 keywordstyle=\color{magenta},
    numberstyle=\tiny\color{codegray},
    stringstyle=\color{codepurple},
    basicstyle=\small\ttfamily,
    breakatwhitespace=false,         
    breaklines=false,                 
    captionpos=b,                    
    keepspaces=false,                 
    numbersep=5pt,                  
    showspaces=false,                
    showstringspaces=false,
    showtabs=false,                  
    tabsize=2,
    escapeinside=``
}

\lstdefinestyle{Cstyle}{
    language=Java,
    aboveskip=3mm,
    showstringspaces=false,
    columns=flexible,
    numbers=left,
    backgroundcolor=\color{backcolour},
    commentstyle=\color{codegreen},
    keywordstyle=\color{magenta},
    numberstyle=\tiny\color{codegray},
    stringstyle=\color{codepurple},
    basicstyle=\scriptsize\ttfamily,
    breakatwhitespace=false,         
    breaklines=true,                 
    captionpos=b,                    
    keepspaces=false,                 
    numbersep=5pt,                  
    showspaces=false,                
    showstringspaces=false,
    showtabs=false,                  
    tabsize=2,
     escapeinside={(*@}{@*)}
}

 \usepackage[preprint]{neurips_2026}

\usepackage[utf8]{inputenc} 
\usepackage[T1]{fontenc}    
\usepackage{hyperref}       
\usepackage{url}            
\usepackage{booktabs}       
\usepackage{amsfonts}       
\usepackage{nicefrac}       
\usepackage{microtype}      
\usepackage{xcolor}         
\usepackage{xcolor}
\usepackage{listings}
\usepackage{tcolorbox}
\usepackage{listings}
\usepackage{caption}
\usepackage{array}

\definecolor{codebg}{RGB}{248,248,248}
\definecolor{codeborder}{RGB}{210,210,210}
\definecolor{lineyellow}{RGB}{255,232,120}
\definecolor{linered}{RGB}{255,130,110}
\definecolor{keywordpurple}{RGB}{150,0,150}
\definecolor{linegreen}{RGB}{190,235,200}

\newcommand{\kw}[1]{\textcolor{keywordpurple}{\textbf{#1}}}
\newcommand{\lnum}[1]{\textcolor{gray}{#1}}
\newcommand{\hly}[1]{%
  \begingroup
  \setlength{\fboxsep}{1pt}%
  \colorbox{lineyellow}{\makebox[\linewidth][l]{#1}}%
  \endgroup
}

\newcommand{\hlr}[1]{%
  \begingroup
  \setlength{\fboxsep}{1pt}%
  \colorbox{linered}{\makebox[\linewidth][l]{#1}}%
  \endgroup
}

\newcommand{\hlg}[1]{%
  \begingroup
  \setlength{\fboxsep}{1pt}%
  \colorbox{linegreen}{\makebox[\linewidth][l]{#1}}%
  \endgroup
}

\lstdefinestyle{javacode}{
    language=Java,
    basicstyle=\ttfamily\small,
    keywordstyle=\color{keywordpurple}\bfseries,
    commentstyle=\color{commentgray},
    numbers=left,
    numberstyle=\scriptsize\color{gray},
    stepnumber=1,
    numbersep=8pt,
    showstringspaces=false,
    breaklines=true,
    keepspaces=true,
    columns=fullflexible,
    tabsize=4,
    escapeinside={(*@}{@*)}
}
\title{{\testgen}: Neuro-Symbolic Guided Test Generation with Large Language Models}

\author{%
 Ruixin Zhang \\
 York University\\
 \texttt{jason666@my.yorku.ca} \\
  \And
  Jiho Shin \\
  York University \\
  \texttt{jihoshin@yorku.ca} \\ 
  \AND
  Hung Viet Pham \\
  York University \\
  \texttt{hvpham@yorku.ca} \\
  \And
  Song Wang \\
  York University \\
  \texttt{wangsong@yorku.ca} \\
}

\begin{document}

\maketitle

\begin{abstract}

Ensuring high structural coverage remains a fundamental challenge in automated test generation, particularly for complex software systems where reaching specific lines or branches requires satisfying intricate control- and data-flow constraints. Large Language Models (LLMs) have recently demonstrated strong capabilities in producing human-like test cases; however, they often struggle to generate inputs that satisfy precise path conditions. 
Conversely, symbolic execution can systematically derive such constraints, but it often fails to construct realistic, executable test cases and is constrained by scalability limitations.

In this paper, we introduce {\testgen}, a hybrid approach that integrates symbolic execution with LLM-driven test synthesis to generate test cases targeting on-demand code coverage. Given a set of target statements within a method, {\testgen} first employs a symbolic analysis engine (i.e., the \textsc{Z3} SMT solver) to extract path-specific constraints and construct a symbolic guidance specification for the desired coverage goal. This specification is then used to guide an LLM in synthesizing concrete test cases that are both structurally valid and semantically meaningful. 
For paths involving complex object-related constraints that are difficult for SMT solvers to handle, {\testgen} leverages LLMs to infer plausible constraints. Furthermore, {\testgen} incorporates an iterative feedback loop that validates LLM-generated tests and provides corrective guidance until the target line or branch is covered or a limit is reached.  
Our empirical evaluation on a widely used benchmark demonstrates that {\testgen} significantly outperforms the state-of-the-art approach across multiple LLMs, including Llama 3.3 70B1, GPT-4o Mini, Claude 3.5 Haiku3, and Claude Sonnet 4.6.  

\end{abstract}

\section{Introduction}
\label{sec:intro}

\textbf{Motivation:} Software testing remains one of the most effective and widely used approaches for revealing faults in modern software systems~\cite{ammann2017introduction}. 
Despite decades of research, automatically generating high–quality tests that achieve precise structural coverage, such as reaching a specific line, branch, or path, remains a fundamental challenge~\cite{schafer2023empirical}. Traditional test generation techniques, such as symbolic execution~\cite{baldoni2018survey}, search-based testing~\cite{formica2023search}, and concolic execution~\cite{yun2018qsym}, have made substantial progress toward systematic exploration of program behavior. However, these techniques often struggle to scale to real-world software and frequently fail to produce readable, maintainable tests that developers can easily adopt within existing test suites.

Recent advancements in Large Language Models (LLMs) have opened new possibilities for automated test generation~\cite{wang2024software}. LLMs demonstrate remarkable ability to synthesize human-like unit tests, construct realistic input objects, compose correct API sequences, and integrate naturally with frameworks such as JUnit~\cite{tufano2020unit,watson2020learning,dinella2022toga}. 
Yet, LLM-based test generation suffers from a critical limitation: while LLMs excel at generating plausible tests, they lack the precise reasoning required to construct tests that satisfy specific path constraints. 
Achieving a target branch or line often requires meeting intricate control-flow and data-flow conditions, maintaining object invariants, or triggering specific program states, capabilities that LLMs alone cannot reliably guarantee~\cite{wu2024natural,wu2026palm}.

To address this limitation, recent approaches have attempted to combine LLMs with program analysis techniques. One state-of-the-art approach is Panta~\cite{gu2025llm}, which integrates static control-flow analysis and dynamic coverage feedback to iteratively guide LLMs toward uncovered program paths and improve structural coverage. While this hybrid design demonstrates the potential of combining LLMs with traditional analysis techniques, it still places excessive reliance on the LLM’s reasoning and code-generation capabilities. In practice, Panta directly includes uncovered execution paths in prompts without determining whether those paths are feasible or satisfiable. As a result, the LLM may repeatedly attempt to generate tests for infeasible paths, leading to wasted token budgets, redundant generations, and unstable performance. Furthermore, because Panta does not explicitly resolve path constraints, infer concrete input conditions, or reason about method signatures and object states before generation, much of the semantic reasoning burden is delegated entirely to the LLM. This over-reliance can easily confuse the model and lead to ineffective or suboptimal test generation, particularly for programs with complex branching logic and intricate input dependencies. 

In contrast, symbolic execution~\cite{baldoni2018survey} is highly effective at reasoning about the logical conditions required to reach targeted program behaviors. Given a target line or branch, symbolic execution can systematically derive path constraints, identify necessary input conditions, detect infeasible paths, and uncover required API invocation sequences. However, symbolic execution alone typically produces low-level constraints or concrete input values rather than complete, readable, and developer-friendly test cases. Furthermore, its practical applicability is often limited by scalability challenges, including path explosion and complex heap reasoning in large object-oriented systems. 

These observations expose a clear gap: Symbolic execution is good at constraint reasoning but poor at producing developer-friendly tests. LLMs generate high-quality tests but struggle to satisfy precise coverage goals.

\textbf{Technique:} To bridge this gap, we propose {\testgen}, a neuro-symbolic test generation approach that leverages the complementary strengths of symbolic reasoning and LLMs to achieve targeted structural coverage on demand. Given a set of target statements within a method, {\testgen} first employs a symbolic analysis engine, i.e., the \textsc{Z3} SMT solver~\cite{de2008z3}, to systematically explore feasible execution paths and extract path-specific constraints required to reach each target. These constraints are then encoded into a symbolic guidance specification that captures the necessary input conditions for achieving the desired coverage goal. This specification serves as structured guidance for an LLM, enabling it to synthesize concrete test cases that not only satisfy the underlying path conditions but also remain syntactically correct, executable, and semantically meaningful within the program context. 
For paths involving complex object-oriented behaviors, such as aliasing, heap structures, or API-specific semantics that are difficult for SMT solvers to precisely model, {\testgen} leverages the reasoning capabilities of LLMs to infer plausible constraints and complete partial specifications. This neuro-symbolic collaboration allows the system to overcome limitations of purely symbolic reasoning while preserving guidance grounded in program semantics. Furthermore, {\testgen} incorporates an iterative feedback loop in which generated test cases are executed and validated against the target coverage objectives. When a test fails to cover the intended line or branch, the system analyzes execution feedback (e.g., constraint violations or runtime behaviors) and provides corrective signals to refine subsequent generations. This process continues until the target is successfully covered or a predefined iteration budget is reached, ensuring both effectiveness and efficiency in test generation.

\textbf{Results:} Our empirical evaluation on the widely used Defects4J benchmark~\cite{just2014defects4j}, including 2,971 methods from 130 classes across 14 projects, demonstrates that {\testgen} consistently improves upon the state-of-the-art test generation approach~\cite{gu2025llm} across multiple LLMs, including Llama 3.3 70B, GPT-4o Mini, Claude 3.5 Haiku, and Claude Sonnet 4.6. 
Moreover, {\testgen} combined with Claude Sonnet 4.6 achieves the best overall performance, reaching 92.91\% line coverage and 88.89\% branch coverage while maintaining high test validity. These results demonstrate that integrating symbolic reasoning with LLM-guided synthesis substantially improves both the effectiveness and robustness of automated unit test generation across LLMs of varying capability and cost.



\section{Problem Formulation
}
\label{sec:2}


\begin{figure}
    \centering
    \includegraphics[width=1\linewidth]{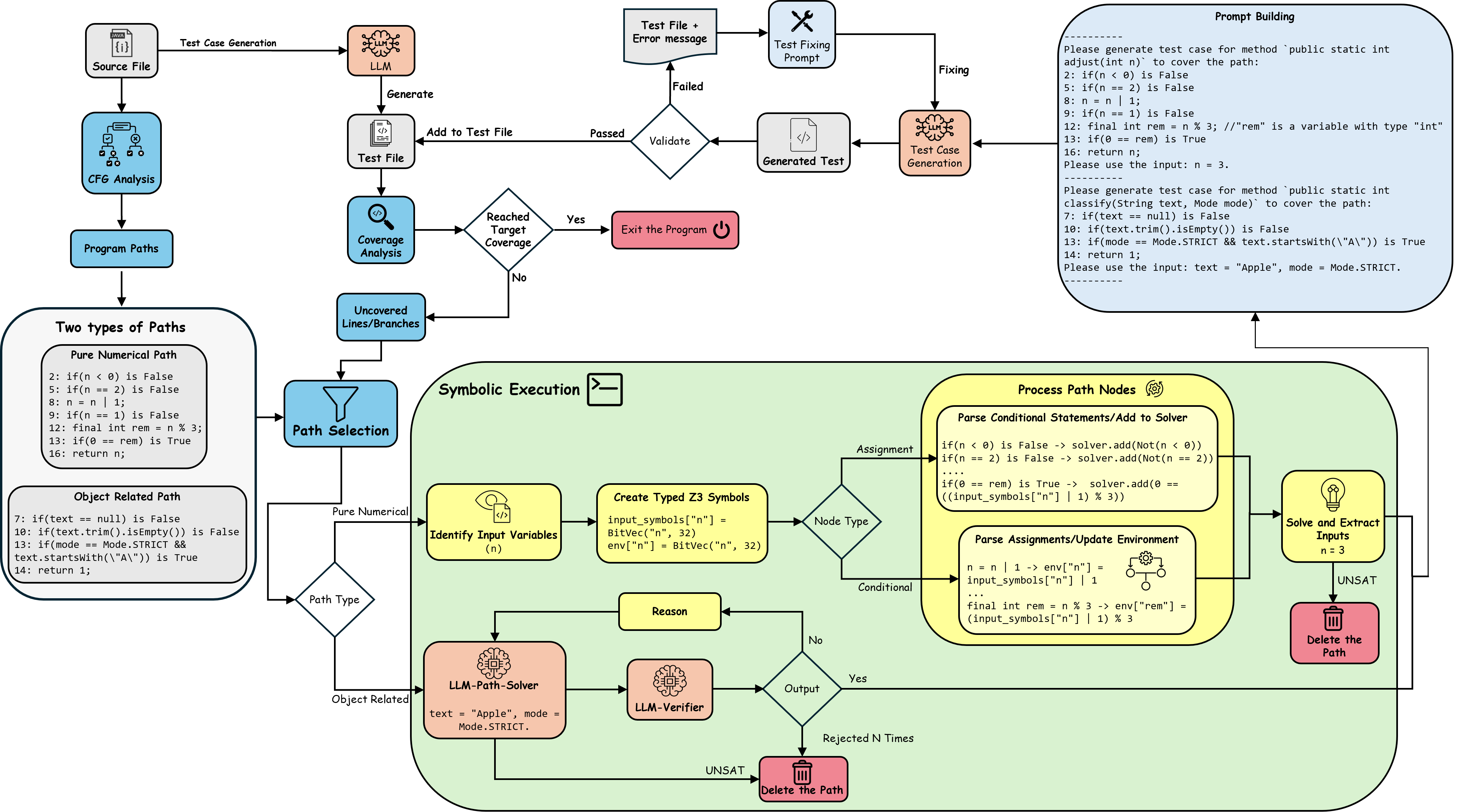}
    \caption{Overview of \testgen
    }
    \label{fig:placeholder}
    
\end{figure}

The overall workflow of our approach is illustrated in Figure~\ref{fig:placeholder}. Given a project and a target source file, we first prompt an LLM to generate an initial set of test cases, providing a baseline for subsequent analysis. We then perform static code analysis to construct the control flow graph (CFG) and enumerate feasible execution paths for each method. 
Then our approach integrates dynamic coverage feedback with CFG-derived paths to identify candidate paths that remain uncovered. 
A path history log is maintained to avoid repeatedly selecting paths that have consistently failed to be covered after multiple attempts. 
An uncovered path is categorized into two types: \textit{pure numerical} and \textit{object-related}. 
A path is considered \textit{pure numerical} when its execution logic consists exclusively of numerical computations, such as arithmetic operations, modulo calculations, comparisons, and bitwise manipulations, and all branch predicates are defined solely over primitive numerical variables without involving object states or external method semantics. 
In contrast, a path is considered \textit{object-related} when its execution depends on object semantics or higher-level program abstractions, including object creation and access, method invocations, library API calls, constant/global object references, or complex string operations. 
These categories are handled differently during the symbolic execution stage, enabling {\testgen} to better address the distinct challenges posed by numerical constraints versus complex object interactions.

\subsection{Pure Symbolic Execution for Solving \textit{Pure Numerical} Paths}

\begin{figure*}[t]
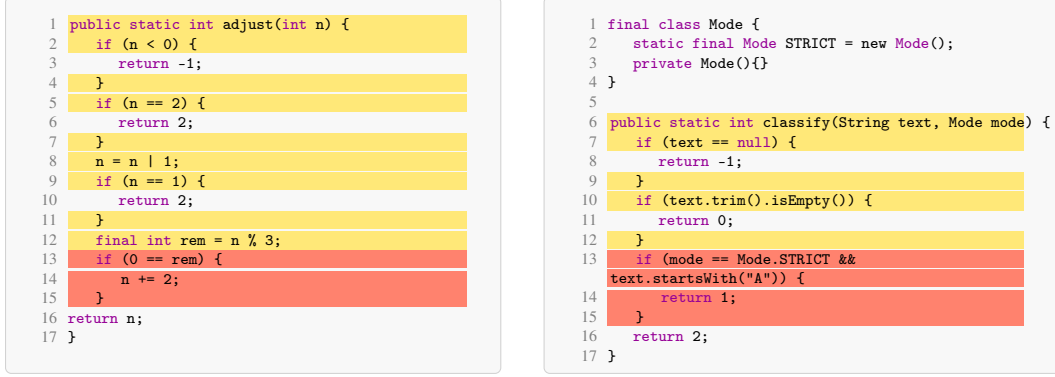

\centering

\begin{subfigure}[t]{0.47\textwidth}
\centering
\begin{tcolorbox}[
    colback=codebg,
    colframe=codeborder,
    boxrule=0.4pt,
    arc=2pt,
    left=2pt,
    right=2pt,
    top=4pt,
    bottom=4pt,
    width=\linewidth,
     height=5cm,
]
\tiny
\setlength{\tabcolsep}{2pt}
\renewcommand{\arraystretch}{1.05}
\begin{tabular}{
    >{\raggedleft\arraybackslash}p{0.07\linewidth}
    p{0.84\linewidth}
}
\lnum{1}  & \texttt{\hly{\kw{public static int} adjust(\kw{int} n) \{}} \\
\lnum{2}  & \texttt{\hly{\hspace*{1.5em}\kw{if} (n < 0) \{}} \\
\lnum{3}  & \texttt{\hspace*{3em}\kw{return} -1;} \\
\lnum{4}  & \texttt{\hly{\hspace*{1.5em}\}}} \\
\lnum{5}  & \texttt{\hly{\hspace*{1.5em}\kw{if} (n == 2) \{}} \\
\lnum{6}  & \texttt{\hspace*{3em}\kw{return} 2;} \\
\lnum{7}  & \texttt{\hly{\hspace*{1.5em}\}}} \\
\lnum{8}  & \texttt{\hly{\hspace*{1.5em}n = n | 1;}} \\
\lnum{9}  & \texttt{\hly{\hspace*{1.5em}\kw{if} (n == 1) \{}} \\
\lnum{10} & \texttt{\hspace*{3em}\kw{return} 2;} \\
\lnum{11} & \texttt{\hly{\hspace*{1.5em}\}}} \\
\lnum{12} & \texttt{\hly{\hspace*{1.5em}\kw{final int} rem = n \% 3;}} \\
\lnum{13} & \texttt{\hlr{\hspace*{1.5em}\kw{if} (0 == rem) \{}} \\
\lnum{14} & \texttt{\hlr{\hspace*{3em}n += 2;}} \\
\lnum{15} & \texttt{\hlr{\hspace*{1.5em}\}}} \\
\lnum{16} & \texttt{{\kw{return} n;}} \\
\lnum{17} & \texttt{{\}}} \\
\end{tabular}
\end{tcolorbox}
\caption{\footnotesize An example of an uncovered \textit{pure numerical} path in method \texttt{adjust()}, where execution starts from the method entry, bypasses the earlier return branches, and reaches the branch \texttt{rem == 0} at lines 13--15, which is not covered by the current test suite.}
\label{fig:adjust_path}
\end{subfigure}
\hfill
\begin{subfigure}[t]{0.49\textwidth}
\centering
\begin{tcolorbox}[
    colback=codebg,
    colframe=codeborder,
    boxrule=0.4pt,
    arc=2pt,
    left=2pt,
    right=2pt,
    top=4pt,
    bottom=4pt,
    width=\linewidth,
        height=5cm,
]
\tiny
\setlength{\tabcolsep}{2pt}
\renewcommand{\arraystretch}{1.05}
\begin{tabular}{
    >{\raggedleft\arraybackslash}p{0.07\linewidth}
    p{0.84\linewidth}
}
\lnum{1}  & \texttt{\kw{final class} Mode \{} \\
\lnum{2}  & \texttt{\hspace*{1.5em}\kw{static final Mode} STRICT = new \kw{Mode}();} \\
\lnum{3}  & \texttt{\hspace*{1.5em}\kw{private} Mode()\{\}} \\
\lnum{4}  & \texttt{\}} \\
\lnum{5}  & \texttt{} \\
\lnum{6}  & \texttt{\hly{\kw{public static int} classify(String text, Mode mode) \{}} \\
\lnum{7}  & \texttt{\hly{\hspace*{1.5em}\kw{if} (text == \kw{null}) \{} }\\
\lnum{8}  & \texttt{\hspace*{3em}\kw{return} -1;} \\
\lnum{9}  & \texttt{\hly{\hspace*{1.5em}\}} }\\
\lnum{10} & \texttt{\hly{\hspace*{1.5em}\kw{if} (text.trim().isEmpty()) \{} }\\
\lnum{11} & \texttt{\hspace*{3em}\kw{return} 0;} \\
\lnum{12} & \texttt{\hly{\hspace*{1.5em}\}}} \\
\lnum{13} & \texttt{\hlr{\hspace*{1.5em}\kw{if} (mode == Mode.STRICT \&\&} \hlr{text.startsWith("A")) \{} } \\
\lnum{14} & \texttt{\hlr{\hspace*{3em}\kw{return} 1;}} \\
\lnum{15} & \texttt{\hlr{\hspace*{1.5em}\}}} \\
\lnum{16} & \texttt{\hspace*{1.5em}\kw{return} 2;} \\
\lnum{17} & \texttt{\}} \\
\end{tabular}
\end{tcolorbox}
\caption{\footnotesize An example of an uncovered \textit{object-related} path in method \texttt{classify()}, where execution reaches the object-dependent branch \texttt{mode == Mode.STRICT \&\& text.startsWith("A")} at lines 13--15, which is not covered by the current test suite.}
\label{fig:classify_path}
\end{subfigure}

\caption{Examples for \textit{pure numerical} path and \textit{object-related} path}
\label{fig:two_path_examples}
\end{figure*}

For \textit{pure numerical} paths, we employ SMT-based symbolic execution to precisely solve path constraints without relying on LLM reasoning.  Figure~\ref{fig:adjust_path} illustrates such a case. In the example, the uncovered branch at lines 13--15 depends entirely on numerical constraints derived from the input variable \texttt{n}. Since the path contains no object states, heap references, or complex API semantics, it can be solved accurately using symbolic reasoning alone. 

To solve this type of path, we leverage the Z3 SMT solver~\cite{de2008z3}. We first extract the method parameters and create corresponding symbolic variables according to their declared types. During symbolic execution, the selected path is traversed sequentially while maintaining a symbolic environment that records the symbolic value of each program variable. Assignment statements update the symbolic state, whereas the original symbolic inputs are preserved for final model construction.
For each branch condition along the path, we translate the predicate into a Z3-compatible constraint. If the execution path follows the false branch of a condition, the predicate is negated before being added to the solver (e.g., we translate \texttt{if (n < 0) is False} to \texttt{Not(n < 0)} and add it to the solver). Numerical expressions involving arithmetic operators, modulo operations, and bitwise operations are directly encoded into SMT formulas. In the example shown in Figure~\ref{fig:adjust_path}, the symbolic executor accumulates constraints from the preceding branches and finally derives the constraint associated with line~13, i.e., \texttt{0 == rem}, where \texttt{rem = n \% 3}. Solving the resulting constraint system enables the solver to synthesize concrete input values that drive execution into the uncovered branch at line~14.
To control solving complexity and maintain efficiency, we currently skip loop bodies during symbolic execution, as loop handling typically requires loop unrolling or invariant inference, both of which can substantially increase constraint complexity and solving overhead. After all constraints along the selected path are collected, Z3 checks the satisfiability of the path condition. If the constraint set is satisfiable, we extract concrete input assignments from the generated model and use them as guidance for subsequent test generation. Otherwise, the path is regarded as infeasible and excluded from the current generation attempt. 


\subsection{LLM-Based Path Solving for \textit{Object-related} Paths}


While symbolic execution is effective for solving numerical constraints, it becomes significantly less effective on \textit{object-related} paths. 
Such paths often depend on semantic knowledge that cannot be easily encoded into SMT constraints. Figure~\ref{fig:classify_path} presents an example of an \textit{object-related} path. In this example, reaching the uncovered branch at lines~13--15 requires simultaneously satisfying two semantic conditions: the input object \texttt{mode} must reference the singleton object \texttt{Mode.STRICT}, and the string parameter \texttt{text} must satisfy the predicate \texttt{text.startsWith("A")}. These conditions involve object identity comparison and string API semantics, which are difficult to model precisely using traditional symbolic execution alone. 

To address this challenge, we employ an LLM-based path-solving strategy using a specialized component called \textit{LLM-Path-Solver}. Given a \textit{object-related} path, we avoid translating the path into SMT constraints. Instead, we provide the path information directly to the \textit{LLM-Path-Solver}, including the source code, method signature, uncovered branch conditions, and execution path context. Based on this information, the model reasons about the program semantics and synthesizes concrete method inputs capable of driving execution toward the target branch. 
For the example shown in Figure~\ref{fig:classify_path}, the \textit{LLM-Path-Solver} infers that the branch condition at line~13 can only be satisfied when \texttt{mode} equals \texttt{Mode.STRICT} and the input string begins with the character \texttt{"A"}. Accordingly, it may generate an input such as \texttt{classify("Apple", Mode.STRICT)}, which enables execution to enter the uncovered branch and cover line~14.

Since LLM-generated inputs may still contain reasoning mistakes or semantic hallucinations, we further introduce a secondary validation stage using another model named \textit{LLM-Verifier}. The verifier symbolically traces the target method using the generated input and checks whether the execution indeed follows the intended path. If the verification fails, the rejection reason, together with the invalid input, is fed back to the \textit{LLM-Path-Solver} for iterative refinement. This feedback-driven process continues until a valid input is generated or a predefined retry limit is reached.  
Once a valid input is confirmed, the generated concrete values are appended to the downstream test-generation prompt as path guidance examples. 

\subsection{Prompt Building and Test Case Generation}

After symbolic execution, the candidate paths and their corresponding input references are formatted into prompts for test generation. 
To enrich the context provided to the LLM, we augment these prompts with explicit type information by annotating variables directly in the code as comments. For example, an initialization such as \texttt{Date date = new Date()} is transformed into \texttt{Date date = new Date() // ``date'' is a variable of type ``Date''}. We similarly handle container types (e.g., arrays, \texttt{ArrayList}, \texttt{Set}, and \texttt{Map}) by appending element-type annotations, such as \texttt{// ``list'' is a List of type ``Integer''} for \texttt{List<Integer> list = new List<>()}. 

Figure~\ref{fig:placeholder} (Prompt Building part) illustrates the template of the test-generation prompt with integrated path-resolving information. The path information combines three types of code: (1) variable definition or object creation code, augmented with variable descriptions; (2) normal operation code, such as calculations and method calls; and (3) conditional or loop code, represented by True/False conditions indicating execution requirements. Finally, we also append an input guide derived from  Z3 SMT solver or LLMs in the prompt. 

The prompt is then fed to the LLM to generate test cases. If a generated test fails (e.g., due to compilation or runtime errors), we provide the error messages along with the failed test file back to the LLM for iterative refinement. Only tests that successfully pass are retained and appended to the existing test suite. The augmented test suite is then re-evaluated for coverage, and the process repeats until the target coverage is achieved or a predefined limit is reached. 
\section{Experiments}
\label{sec:exp}

\begin{wrapfigure}{r}{0.6\textwidth}

\centering 
\caption{Experiment project statistics
}
\label{tab:dataset}
\resizebox{0.6\textwidth}{!}{%
\begin{tabular}{l l r r|cc|cc}
\toprule
Identifier & Project & \#Class & \#Methods 
& \multicolumn{2}{c|}{Pure Numerical Path} 
& \multicolumn{2}{c}{Object-Related Path} \\
\cmidrule(lr){5-6} \cmidrule(lr){7-8}
& & & 
& Count & Ratio (\%) 
& Count & Ratio (\%) \\
\midrule
Cli        & Cli-40f              & 2  & 31  & 6   & 11.54 & 46   & 88.46 \\
Codec      & Codec-18f            & 7  & 78  & 3   & 1.62  & 182  & 98.38 \\
Collections& Collections-28f      & 5  & 218 & 15  & 3.62  & 399  & 96.38 \\
Compress   & Compress-47f         & 9  & 65  & 42  & 13.68 & 265  & 86.32 \\
Csv        & Csv-16f              & 3  & 72  & 21  & 18.58 & 92   & 81.42 \\
Gson       & Gson-16f             & 4  & 75  & 53  & 17.15 & 256  & 82.85 \\
JCore      & JacksonCore-26f      & 9  & 161 & 110 & 23.26 & 363  & 76.74 \\
JDatabind  & JacksonDatabind-112f & 9  & 371 & 251 & 25.18 & 746  & 74.82 \\
JXml       & JacksonXml-5f        & 4  & 118 & 26  & 5.83  & 420  & 94.17 \\
Jsoup      & Jsoup-93f            & 8  & 171 & 74  & 22.09 & 261  & 77.91 \\
JXPath     & JXPath-22f           & 12 & 115 & 49  & 7.95  & 567  & 92.05 \\
Lang       & Lang-4f              & 17 & 586 & 273 & 24.59 & 837  & 75.41 \\
Math       & Math-2f              & 30 & 452 & 346 & 22.97 & 1,160 & 77.03 \\
Time       & Time-13f             & 11 & 458 & 141 & 25.97 & 402  & 74.03 \\
\midrule
Total/Ave  & ---                  & 130 & 2,971 
& 1,410 & 19.04 & 5,996 & 80.96 \\
\bottomrule
\end{tabular}
}
\end{wrapfigure}

\subsection{Datasets}

To ensure a fair comparison and following the state-of-the-art setting, we use the same dataset as \cite{gu2025llm}, which includes 14 real-world subjects from Defects4J~\cite{just2014defects4j}. The number of classes ranges from 2 to 30, and the number of methods under test ranges from 31 to 586. Table~\ref{tab:dataset} presents the selected subjects. The column Project denotes the project name together with its Defects4J version identifier. In total, we evaluate 130 classes containing 2,971 methods under test. 
Across all projects, \textit{object-related} paths overwhelmingly dominate execution, accounting for 80.96\% of all paths on average, compared to only 19.04\% for \textit{pure numerical} paths. This trend is consistent across nearly all subjects, indicating that real-world Java programs are largely driven by object interactions rather than purely numerical computations.

For a controlled and comparable evaluation, we discard the original test suites in each project and generate tests from scratch for all target classes, which helps eliminate any potential bias introduced by existing tests and ensures that all approaches are evaluated under the same conditions.

\subsection{Baseline and LLM Selection}

To evaluate {\testgen}, we compare it against the state-of-the-art LLM-based test generation technique, \textit{Panta}~\cite{gu2025llm}, which combines static control-flow analysis and dynamic coverage feedback to iteratively guide LLMs toward generating tests for uncovered execution paths.

For the experimental LLM configuration, we follow \textit{Panta}~\cite{gu2025llm} and adopt the same evaluation protocol, benchmark setup, and language models to ensure a fair and consistent comparison. 
Specifically, we experiment with the same four representative large language models used in~\cite{gu2025llm}, including Meta’s Llama 3.3 70B\footnote{\url{https://www.llama.com/docs/model-cards-and-prompt-formats/llama3_3/}}, 
OpenAI’s GPT-4o Mini\footnote{\url{https://developers.openai.com/api/docs/models/gpt-4o-mini}}, and Anthropic’s Claude 3.5 Haiku\footnote{\url{https://www.anthropic.com/news/3-5-models-and-computer-use}}. In addition, we include a strong state-of-the-art coding-oriented model, Claude Sonnet 4.6\footnote{\url{https://www.anthropic.com/news/claude-sonnet-4-6}}, to better contextualize the performance of our approach.

All evaluated LLMs support a context window of up to 128K tokens, ensuring sufficient capacity for handling large code contexts and specifications. To maintain consistency and eliminate confounding factors, we fix the decoding parameters across all models: the maximum number of generated tokens is set to 4096, and the temperature is set to 0.2. This relatively low temperature encourages stable and near-deterministic outputs while still allowing limited variability.

\subsection{Evaluation Metrics}

We adopt widely used quality metrics from the test generation literature~\cite{gu2025llm,chen2024chatunitest,schafer2023empirical}, including \textbf{Line Coverage}, \textbf{Branch Coverage}, \textbf{Pass Rate}, and \textbf{Cost}. 

\textbf{Line Coverage.} measures the proportion of executable source code lines that are exercised by the generated test suite. It is computed as the number of lines executed at least once divided by the total number of executable lines in the target class. 
\textbf{Branch Coverage.} measures the proportion of control-flow branches (e.g., true/false outcomes of conditional statements such as \texttt{if}, \texttt{switch}, and loop conditions) that are executed by the test suite. It is calculated as the number of covered branches divided by the total number of branches in the class. 
\textbf{Pass Rate.} The pass rate is defined as the percentage of generated tests that execute successfully when incorporated into the test suite. A test is considered passing if it compiles successfully and runs without failures or runtime exceptions. 
\textbf{Cost}. Since different LLMs incur varying usage expenses, we report the corresponding monetary \textbf{cost} of {\testgen} with each model. 


As our evaluation is conducted at the class level rather than the method level, we compute line and branch coverage for each class individually and report the average coverage across all target classes within each project.

\section{Results}
\label{sec:results}

\subsection{Main Results of {\testgen}'s Performance}

\begin{table}[t]
\centering
\caption{Comparison of {\tool} and Panta~\cite{gu2025llm} across projects with two base models, i.e., Claude 3.5 Haiku and Claude 4.6 Sonnet.}

\label{tab:panta_testgen}

\resizebox{\textwidth}{!}{  
\begin{tabular}{l|cc|cc|cc|cc|cc|cc}
\toprule
\multirow{3}{*}{Project} 
& \multicolumn{4}{c|}{Line Coverage} 
& \multicolumn{4}{c|}{Branch Coverage} 
& \multicolumn{4}{c}{Pass Rate} \\

\cmidrule(lr){2-5} \cmidrule(lr){6-9} \cmidrule(lr){10-13}

& \multicolumn{2}{c|}{Panta~\cite{gu2025llm}} 
& \multicolumn{2}{c|}{{\tool}} 
& \multicolumn{2}{c|}{Panta~\cite{gu2025llm}} 
& \multicolumn{2}{c|}{{\tool}} 
& \multicolumn{2}{c|}{Panta~\cite{gu2025llm}} 
& \multicolumn{2}{c}{{\tool}} \\

\cmidrule(lr){2-3} \cmidrule(lr){4-5}
\cmidrule(lr){6-7} \cmidrule(lr){8-9}
\cmidrule(lr){10-11} \cmidrule(lr){12-13}

& Claude 3.5 & Claude 4.6 & Claude 3.5 & Claude 4.6
&Claude 3.5 & Claude 4.6 & Claude 3.5 & Claude 4.6
& Claude 3.5 & Claude 4.6 & Claude 3.5 & Claude 4.6 \\
\midrule

Cli        & 91.14 & 98.82 & 91.14 & 98.82 & 82.74 & 92.08 & 83.15 & 92.49 & 87.31 & 97.18 & 56.28 & 98.02 \\
Codec      & 83.30 & 98.45 & 83.91 & 98.45 & 79.40 & 94.86 & 80.70 & 95.03 & 64.98 & 98.20 & 46.97 & 98.15\\
Collections& 80.35 & 98.97 & 82.14 & 99.24 & 80.59 & 96.36 & 81.79 & 96.79 & 85.63 & 99.78 & 84.84 & 99.56\\
Compress   & 59.69 & 91.91 & 62.80 & 92.76 & 51.28 & 85.04 & 57.57 & 86.79 & 51.80 & 90.29 & 58.44 & 90.40\\
Csv        & 58.11 & 83.80  & 74.90 & 86.19 & 46.42 & 73.56 & 62.44 & 77.14 & 71.83 & 98.27 & 71.81 & 98.31\\
Gson       & 85.19 & 92.05 & 88.01 & 92.05 & 74.25 & 88.52 & 80.06 & 88.52 & 60.94 & 99.08 & 54.65 & 99.18\\
JCore      & 73.88 & 87.70  & 76.32 & 89.06 & 67.01 & 85.94 & 69.84 & 87.42 & 69.31 & 97.38 & 77.55 & 97.21\\
JDatabind  & 35.00 & 86.95 & 51.22 & 88.34 & 34.45 & 82.17 & 47.69 & 83.86 & 47.26 & 99.35 & 59.19 & 99.02\\
JXml       & 46.42 & 77.20 & 54.57 & 77.20 & 50.34 & 79.70 & 58.47 & 79.70 & 44.23 & 90.87 & 71.10 & 91.15\\
Jsoup      & 86.75 & 96.05 & 88.36 & 96.90 & 73.38 & 88.41 & 79.08 & 90.12 & 81.69 & 98.07 & 73.50 & 99.15\\
JXPath     & 55.31 & 84.84 & 57.17 & 93.74 & 50.46 & 78.53 & 53.73 & 88.36 & 38.44 & 89.13 & 43.30 & 89.35\\
Lang       & 78.17 & 93.23 & 80.02 & 93.87 & 74.13 & 89.36 & 76.93 & 89.95 & 75.44 & 99.90 & 78.57 & 99.20\\
Math       & 80.77 & 95.87 & 84.98 & 96.37 & 73.60 & 92.42 & 78.71 & 93.37 & 76.49 & 98.89 & 77.99 & 99.26\\
Time       & 78.21 & 97.66 & 86.69 & 97.79 & 74.00 & 94.38 & 82.01 & 94.87 & 67.50 & 98.98 & 74.54 & 98.27\\
\hline
Total/Ave  & 70.88 & 91.68 & 75.87 & 92.91 & 65.15 & 87.24 & 70.87 & 88.89 & 65.92 & 96.81 & 66.34 & 96.88\\

\bottomrule
\end{tabular}
}
\end{table}

Panta~\cite{gu2025llm} reported its best performance using Claude-3.5-Haiku; therefore, to ensure a fair comparison with Panta, we adopt Claude-3.5-Haiku as one of our primary evaluation models and use the original results reported in their paper. In addition, we include the latest state-of-the-art coding-oriented model, Claude Sonnet 4.6, to further evaluate the effectiveness of {\testgen} under a stronger LLM setting.

Our main results are presented in Table~\ref{tab:panta_testgen}, which compares Panta~\cite{gu2025llm} and {\testgen} using both Claude-3.5-Haiku and Claude Sonnet 4.6. 
Overall, {\testgen} consistently outperforms Panta across most projects and metrics, achieving higher average line coverage, branch coverage, and pass rates under both LLM settings. These results demonstrate the effectiveness of combining symbolic execution with LLM-guided test synthesis. 
For line coverage, {\testgen} improves the average coverage from 70.88\% to 75.87\% with Claude 3.5 and from 91.68\% to 92.91\% with Claude 4.6, with notable gains on projects such as \textit{Csv}, \textit{Jsoup}, and \textit{JXPath}. 
For branch coverage, {\testgen} increases the average coverage from 65.15\% to 70.87\% using Claude 3.5 and from 87.24\% to 88.89\% using Claude 4.6, demonstrating a stronger ability to cover difficult execution paths. 
{\testgen} also maintains high test validity, slightly improving the average pass rate from 65.92\% to 66.34\% with Claude 3.5 and from 96.81\% to 96.88\% with Claude 4.6.

{\testgen} achieves larger improvements over Panta when using earlier-generation LLMs such as Claude 3.5 compared to stronger models such as Claude 4.6. This suggests that the symbolic reasoning and constraint-solving capabilities introduced by {\testgen} can effectively compensate for the weaker reasoning abilities of smaller or less advanced LLMs.

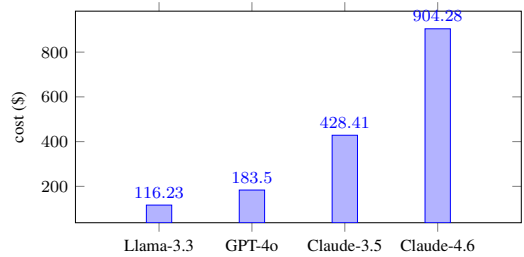
\begin{wrapfigure}{r}{0.5\textwidth}
  \centering
  \resizebox{0.5\textwidth}{!}{
\begin{tikzpicture}
\begin{axis}[
font=\footnotesize,
    ybar,
    bar width=12pt,
    width=9cm,
    height=5.1cm,
    enlarge x limits=0.3,
    legend style={at={(0.5,-0.25)}, anchor=north, legend columns=-1},
    ylabel={cost (\$)},
    symbolic x coords={Llama-3.3, GPT-4o, Claude-3.5, Claude-4.6},
    xtick=data,
    nodes near coords,
    nodes near coords align={vertical},
]

\addplot coordinates {(Llama-3.3, 116.23) (GPT-4o, 183.50) (Claude-3.5, 428.41) (Claude-4.6, 904.28)};


\end{axis}
\end{tikzpicture}
}
\caption{\small Comparison of cost across experimental LLMs ( Llama 3.3 70B, GPT-4o Mini, Claude 3.5 Haiku, and Claude Sonnet 4.6).
}
\label{fig:llm_cost_tokens}

\end{wrapfigure}

\subsection{Impact of Different Models}

\begin{table}[t]
\centering
\caption{Performance of {\tool} with different LLMs} 
\label{tab:llm_comparison}
 \resizebox{\textwidth}{!}{  

\begin{tabular}{l|cccc|cccc|cccc}
\toprule
\multirow{2}{*}{Project} 
& \multicolumn{4}{c|}{Line Coverage} 
& \multicolumn{4}{c|}{Branch Coverage} 
& \multicolumn{4}{c}{Pass Rate} \\
\cmidrule(lr){2-5} \cmidrule(lr){6-9} \cmidrule(lr){10-13}
& Llama 3.3 & GPT4o & Claude 3.5& Claude 4.6 
& Llama 3.3 & GPT4o & Claude 3.5& Claude 4.6  
& Llama 3.3 & GPT4o & Claude 3.5& Claude 4.6  \\
\midrule
Cli        & 83.96 & 89.04 & 91.14 & 98.82 & 70.27 & 70.43 & 83.15 & 92.49 & 35.33 & 47.97 & 56.28 & 98.02 \\
Codec      & 75.01 & 57.07 & 83.91 & 98.45 & 67.87 & 49.4  & 80.70 & 95.03 & 45.99 & 31.58 & 46.97 & 98.15 \\
Collections& 84.7  & 66.69 & 82.14 & 99.24 & 83.6  & 62.12 & 81.79 & 96.79 & 59.60 & 49.35 & 84.84 & 99.56 \\
Compress   & 51.17 & 42.3  & 62.80 & 92.76 & 45.75 & 34.38 & 57.57 & 86.79 & 31.22 & 18.85 & 58.44 & 90.40 \\
Csv        & 70.19 & 71.09 & 74.90 & 86.19 & 52.63 & 55.42 & 62.44 & 77.14 & 34.29 & 22.71 & 71.81 & 98.31 \\
Gson       & 76.14 & 80.46 & 88.01 & 92.05 & 62.88 & 69.78 & 80.06 & 88.52 & 35.58 & 69.15 & 54.65 & 99.18 \\
JCore      & 60.08 & 55.28 & 76.32 & 89.06 & 54.65 & 48.95 & 69.84 & 87.42 & 44.50 & 50.80 & 77.55 & 97.21 \\
JDatabind  & 57.45 & 37.33 & 51.22 & 88.34 & 50.67 & 33.67 & 47.69 & 83.86 & 30.09 & 54.28 & 59.19 & 99.02 \\
JXml       & 50.24 & 38.69 & 54.57 & 77.20 & 40.98 & 35.53 & 58.47 & 79.70 & 43.04 & 32.57 & 71.10 & 91.15 \\
Jsoup      & 86.83 & 73.67 & 88.36 & 96.9  & 69.72 & 57.78 & 79.08 & 90.12 & 43.38 & 45.23 & 73.50 & 99.15 \\
JXPath     & 54.72 & 41.41 & 57.17 & 93.74 & 47.67 & 35.98 & 53.73 & 88.36 & 35.19 & 32.71 & 43.30 & 89.35 \\
Lang       & 80.29 & 73.12 & 80.02 & 93.87 & 71.27 & 65.52 & 76.93 & 89.95 & 41.10 & 50.17 & 78.57 & 99.20 \\
Math       & 66.9  & 55.56 & 84.98 & 96.37 & 59.41 & 49.42 & 78.71 & 93.37 & 49.01 & 33.90 & 77.99 & 99.26 \\
Time       & 74.08 & 87.39 & 86.69 & 97.79 & 65.68 & 76.73 & 82.01 & 94.87 & 39.51 & 41.21 & 74.54 & 98.27 \\\hline
Total/Ave  & 69.41 & 62.08 & 75.87 & 92.91 & 60.22 & 53.22 & 70.87 & 88.89 & 40.56 & 41.46 & 66.34 & 96.88\\
\bottomrule
\end{tabular}
}
\end{table}
We also examine the performance of {\testgen} across different LLMs to evaluate its effectiveness and generalizability under models with varying capabilities and costs. 

Table~\ref{tab:llm_comparison} presents a detailed comparison of {\testgen} when instantiated with different LLMs, namely Llama 3.3 70B, GPT-4o Mini, Claude 3.5 Haiku, and Claude Sonnet 4.6. Overall, the performance varies significantly across models, highlighting the strong impact of the underlying LLM on test generation effectiveness. 

Among the evaluated models, Claude Sonnet 4.6 achieves the highest average line coverage (92.91\%), branch coverage (88.89\%), and pass rate (96.88\%),  demonstrating the benefit of stronger code reasoning and generation capabilities. Claude 3.5 Haiku also performs competitively, reaching 75.87\% line coverage and 70.87\% branch coverage, indicating that {\testgen} can effectively leverage mid-sized LLMs to achieve strong testing performance.  
In contrast, Llama 3.3 and GPT-4o Mini exhibit noticeably lower performance, particularly in pass rate and branch coverage, suggesting that weaker LLMs struggle more with generating valid tests and satisfying complex execution constraints even with symbolic guidance. Nevertheless, {\testgen} still enables these smaller models to achieve reasonable coverage across multiple projects, highlighting the robustness of the hybrid symbolic execution and LLM-guided framework. 

Figure~\ref{fig:llm_cost_tokens} compares the generation cost of different LLMs. Llama 3.3 70B incurs the lowest cost (\$116.23), followed by GPT-4o-Mini (\$183.50) and Claude 3.5 Haiku (\$428.41), while Claude Sonnet 4.6 is the most expensive at \$904.28. Although Claude Sonnet 4.6 achieves the best coverage and pass rates, these gains come at a significant cost, requiring approximately 7.8$\times$ higher cost than Llama 3.3 70B and 4.9$\times$ higher cost than GPT-4o-Mini. In contrast, Claude 3.5 Haiku offers a more favorable balance between performance and cost, achieving competitive test generation quality at substantially lower expense.

\subsection{Ablation Study}

\begin{table}[t]
\centering
\caption{Ablation Study ({\testgen} with Symbolic Only vs. Full {\testgen}) with two LLMs, i.e., Claude 3.5 Haiku and Claude Sonnet 4.6.}
\label{tab:ablation}
\resizebox{0.9\textwidth}{!}{  
\begin{tabular}{l|cc|cc|cc|cc}
\toprule
\multirow{2}{*}{Project} 
& \multicolumn{4}{c|}{Line Coverage (\%)} 
& \multicolumn{4}{c}{Branch Coverage (\%)} \\
\cmidrule(lr){2-5} \cmidrule(lr){6-9}
& \multicolumn{2}{c|}{Claude 3.5} 
& \multicolumn{2}{c|}{Claude 4.6} 
& \multicolumn{2}{c|}{Claude 3.5} 
& \multicolumn{2}{c}{Claude 4.6}\\
\cmidrule(lr){2-3} \cmidrule(lr){4-5}  \cmidrule(lr){6-7}  \cmidrule(lr){8-9}
& Symbolic Only  & Full  & Symbolic Only  & Full
& Symbolic Only  & Full  & Symbolic Only  & Full \\
\midrule
Cli         & 94.76 & 91.14 & 98.82 & 98.82 & 84.67 & 83.15 & 92.49 & 92.49 \\
Codec       & 83.30 & 83.91 & 98.42 & 98.45 & 79.72 & 80.70 & 95.02 & 95.03 \\
Collections & 81.33 & 82.14 & 99.15 & 99.24 & 81.58 & 81.79 & 96.65 & 96.79 \\
Compress    & 60.76 & 62.80 & 92.35 & 92.76 & 52.52 & 57.57 & 86.40 & 86.79 \\
Csv         & 73.58 & 74.90 & 85.42 & 86.19 & 62.90 & 62.44 & 75.60 & 77.14 \\
Gson        & 86.10 & 88.01 & 92.05 & 92.05 & 75.78 & 80.06 & 88.52 & 88.52 \\
JCore       & 75.59 & 76.32 & 87.72 & 89.06 & 70.66 & 69.84 & 85.98 & 87.42 \\
JDatabind   & 45.09 & 51.22 & 87.47 & 88.34 & 43.68 & 47.69 & 82.87 & 83.86 \\
JXml        & 46.53 & 54.57 & 77.20 & 77.20 & 50.58 & 58.47 & 79.88 & 79.70 \\
Jsoup       & 88.83 & 88.36 & 96.05 & 96.90 & 77.83 & 79.08 & 88.82 & 90.12 \\
JXPath      & 56.63 & 57.17 & 90.79 & 93.74 & 52.72 & 53.73 & 84.67 & 88.36 \\
Lang        & 79.71 & 80.02 & 93.51 & 93.87 & 76.17 & 76.93 & 89.77 & 89.95 \\
Math        & 82.21 & 84.98 & 96.24 & 96.37 & 75.93 & 78.71 & 93.05 & 93.37 \\
Time        & 86.58 & 86.69 & 97.66 & 97.79 & 81.91 & 82.01 & 94.38 & 94.87 \\
\hline
Total/Ave   & 73.45 & 75.87 & 92.35 & 92.91 & 68.53 & 70.87 & 88.15 & 88.89 \\
\bottomrule
\end{tabular}
}
\end{table}

As discussed in Section~\ref{sec:2}, execution paths include both \textit{pure numerical} and \textit{object-related} paths. While \textit{pure numerical} paths can be effectively handled by symbolic execution, \textit{object-related} paths are more challenging and require LLM-based support. In this section, we compare {\testgen} with \textbf{symbolic execution only} against the full {\testgen} framework to evaluate the contribution of the LLM component in path solving. 

Table~\ref{tab:ablation} presents the ablation study comparing the symbolic-only variant of {\testgen} with the full framework under Claude 3.5 Haiku and Claude Sonnet 4.6. Overall, integrating LLM-guided test synthesis consistently improves both line and branch coverage over symbolic execution alone. 
With Claude 3.5, the full framework improves average line coverage from 73.45\% to 75.87\% and branch coverage from 68.53\% to 70.87\%. Similarly, under Claude Sonnet 4.6, the full framework increases average line coverage from 92.35\% to 92.91\% and branch coverage from 88.15\% to 88.89\%. 
The relatively small improvement of the full {\testgen} framework over the symbolic-only variant is partly due to the strong inherent capabilities of modern LLMs. Even without explicitly solving object-related constraints, LLMs can often infer reasonable object constructions, API sequences, and input values from learned programming patterns, allowing the symbolic-only variant to achieve relatively strong performance.

Nevertheless, the full {\testgen} framework consistently performs better, especially on projects with complex object-oriented logic and API interactions. These results highlight the complementary strengths of symbolic execution and LLM-based synthesis: symbolic execution provides precise path reasoning, while LLMs contribute semantic understanding and flexible test generation.
\section{Related Work}
\label{sec:background}

\subsection{LLM-Based Test Generation}

Automated unit test generation using LLMs has been widely researched in recent years. 
AthenaTest \cite{tufano2020unit} used transformer models trained on the focal method context to generate unit tests, and ATLAS \cite{watson2020learning} proposed neural methods that predict meaningful assertions to strengthen fault detection. Later, TOGA \cite{dinella2022toga} shows that machine learning models can learn assertion patterns from existing tests and documentation and successfully generate useful oracles. 
CodaMosa \cite{lemieux2023codamosa} invokes a model to escape coverage plateaus in search-based generation, while Cedar \cite{nashid2023retrieval} retrieves effective demonstrations to stabilize few-shot prompting for assertion-related tasks. Shin et al. \cite{shin2024retrieval} further study retrieval augmented test generation and show that providing LLMs with external domain knowledge can improve unit test generation for library APIs. 
ChatUniTest \cite{chen2024chatunitest} introduces an iterative validation and repair mechanism to fix failed tests, and MuTAP \cite{dakhel2024effective} incorporates feedback from mutation tests to improve the effectiveness of test cases. More recently, Steenhoek et. al. \cite{steenhoek2025reinforcement} proposed a method that used reinforcement learning from automatic quality metrics that helps to reduce test smells. Panta \cite{gu2025llm} used a hybrid approach that increased line and branch coverage by using static program analysis to reveal uncovered paths, which gives LLM more information, and ASTER \cite{pan2025aster} introduced a generic, multi-language unit test generation pipeline that combines program analysis with environment mocking to handle real-world software systems. Together, these studies indicate a switch from one-shot prompting toward iterative and feedback-guided test generation that improves both coverage and test effectiveness.

\subsection{Symbolic Execution}

Symbolic execution is a white box program analysis technique that runs a program using symbolic values in place of concrete inputs and accumulates a path condition that captures the constraints required to follow each explored execution path \cite{king1976symbolic}. By solving these constraints, it can get concrete inputs that lead the program execution toward specific branches and behaviors, making it helpful for systematic test case generation \cite{king1976symbolic}. Early work \cite{king1976symbolic} established the core idea of symbolic execution for program testing, while concolic execution, such as CUTE \cite{sen2005cute}, combined concrete runs with symbolic reasoning to iteratively negate branch predicates and generate new tests with higher path diversity. Scalable engines such as KLEE \cite{cadar2008klee} further demonstrated that symbolic execution can automatically generate high-coverage tests for complex systems, but also highlighted challenges, including path explosion and constraint-solver bottlenecks.

Recent research revisits these challenges by integrating LLMs with symbolic and concolic execution to improve test generation  \cite{wu2025generating2,li2025large,wang2024python,ryan2024code,tu2025cottontail,wang2025can}. 
PALM intentionally avoids SMT-based constraint solving by translating paths into executable variants interpretable by LLMs~\cite{wu2025generating2}.  
Wang et al.~\cite{wang2024python} proposed LLM-Sym to translate complex Python path constraints into Z3 code. 
SymPrompt \cite{ryan2024code} studies coverage-guided regression test generation and proposes a multi-stage prompting strategy that aligns generation with execution paths while feeding type and dependency contexts, thereby helping the model produce executable tests and escape coverage plateaus in practice. Cottontail \cite{tu2025cottontail} employs LLM-driven solving into concolic execution to generate highly structured inputs for programs to avoid wasted testing efforts.  

Different from prior studies, our approach combines precise symbolic reasoning with LLM-based semantic understanding in a unified path-solving framework. By adaptively applying SMT-based symbolic execution to numerically constrained paths and verifier-guided LLM reasoning to semantically complex paths, our method improves both the accuracy and effectiveness of solving uncovered execution paths in real-world test generation scenarios.
\section{Conclusion}

In this paper, we propose {\testgen}, a hybrid approach that combines symbolic execution with LLM-driven test synthesis to generate test cases. Our evaluation on a standard benchmark demonstrates that {\testgen} significantly outperforms the state-of-the-art approach. 
Despite these promising results, {\testgen} still has several limitations. First, the effectiveness of symbolic execution remains constrained by scalability challenges such as path explosion and incomplete modeling of complex heap states. Second, although LLM guidance improves object construction and API usage generation, the approach may still struggle with highly domain-specific libraries or programs requiring deep semantic understanding. Finally, the iterative refinement process introduces additional computational overhead due to repeated compilation, execution, and prompting cycles. These limitations highlight opportunities for future work on improving symbolic scalability, enhancing semantic reasoning capabilities, and reducing generation cost.

\vspace{4pt}
\textbf{Data Availability Statement}: We provide our dataset and source code at \url{{https://anonymous.4open.science/r/Symbolic_Execution_Based_Test_Generation-F4F3/}}.

\bibliographystyle{plain}
\bibliography{ref}




\appendix
\section*{Appendix}

\section{Failure Analysis}

We conduct a qualitative case study to investigate why {\testgen} fails to achieve 100\% coverage, even when provided with concrete input guidance derived from symbolic execution or LLMs. Specifically, we collect all methods for which {\testgen} achieves less than full coverage and perform a manual inspection of the uncovered lines and branches. Using an open coding methodology, we systematically analyze these cases to identify recurring patterns and underlying causes. Our analysis reveals three primary categories of limitations:

\subsection{Partially Covered Lines and Branches} 

In this paper, consistent with prior work, we adopt JaCoCo\footnote{\url{https://www.eclemma.org/jacoco/}}
 as our coverage measurement tool. According to JaCoCo’s coverage model, coverage is reported at multiple granularities (e.g., instruction, line, and branch), and importantly, a branch can be marked as partially covered. Consequently, a method may still be reported as fully covered at the method level even when some of its constituent lines or branches are only partially exercised.

This nuance introduces a subtle but important limitation for coverage-guided approaches. In our setting, prompts are constructed based on uncovered program elements identified by JaCoCo. However, partially covered conditional statements are not always explicitly surfaced or prioritized in this process. As a result, such statements may be omitted from the generated prompts, even though they contain unexplored execution paths. 
In practice, this leads to scenarios in which all methods are reported as fully covered, whereas certain lines, particularly those involving complex conditionals, remain only partially exercised. This discrepancy highlights a gap between reported coverage metrics and the actual thoroughness of path exploration, which can mislead both automated systems and developers relying on these signals for test generation. 


\definecolor{covgreen}{RGB}{198,239,206}   
\definecolor{covyellow}{RGB}{255,235,156}  
\definecolor{covred}{RGB}{255,199,206}     








\begin{figure}[h!]
\centering
\begin{tcolorbox}[
    colback=codebg,
    colframe=codeborder,
    boxrule=0.4pt,
    arc=2pt,
    left=2pt,
    right=2pt,
    top=4pt,
    bottom=4pt,
    width=\linewidth
]
\footnotesize
\setlength{\tabcolsep}{3pt}
\renewcommand{\arraystretch}{1.08}
\begin{tabular}{
    >{\raggedleft\arraybackslash}p{0.04\linewidth}
    p{0.88\linewidth}
}
\lnum{1}  & \texttt{\kw{private int} handleX(\kw{final} String value, \kw{final} DoubleMetaphoneResult result, \kw{int} index) \{} \\
\lnum{2}  & \texttt{\hlg{\hspace*{1.5em}\kw{if} (index == 0) \{}} \\
\lnum{3}  & \texttt{\hlg{\hspace*{3em}result.append('S');}} \\
\lnum{4}  & \texttt{\hlg{\hspace*{3em}index++;}} \\
\lnum{5}  & \texttt{\hspace*{1.5em}\} \kw{else} \{} \\
\lnum{6} & \texttt{\hlg{\hspace*{3em}\kw{if} (!((index == value.length() - 1) \&\&}} \\
\lnum{7} & \texttt{\hly{\hspace*{3em}(contains(value, index - 3, 3, "IAU", "EAU") ||}} \\
\lnum{8} & \texttt{\hlr{\hspace*{3em}contains(value, index - 2, 2, "AU", "OU")))) \{}} \\
\lnum{9} & \texttt{\hspace*{3em}// French e.g. breaux} \\
\lnum{10} & \texttt{\hlg{\hspace*{3em}result.append("KS");}} \\
\lnum{11} & \texttt{\hspace*{3em}\}} \\
\lnum{12} & \texttt{\hlg{\hspace*{3em}index = contains(value, index + 1, 1, "C", "X") ? index + 2 : index + 1;}} \\
\lnum{13} & \texttt{\hspace*{1.5em}\}} \\
\lnum{14} & \texttt{\hlg{\hspace*{1.5em}\kw{return} index;}} \\
\lnum{15} & \texttt{\}} \\
\end{tabular}
\end{tcolorbox}
\caption{JaCoCo-style visualization of coverage for method \texttt{handleX()}. Green indicates fully covered lines, yellow indicates partially covered branches, and red highlights uncovered conditions within a compound predicate.}
\label{fig:jacoco-colored}
\end{figure}

For example, Figure~\ref{fig:jacoco-colored} illustrates the JaCoCo coverage report for the method \texttt{handleX()} in \texttt{DoubleMetaphoneTest.java} from the Codec-18f project, which achieves 99.04\% line coverage and 93.15\% branch coverage. The report indicates that the compound condition between lines 6 and 8 is covered, as the corresponding condition has been evaluated to true during execution. However, this branch is only partially covered, as not all logical combinations of its sub-conditions have been exercised. 
In particular, the LLM fails to generate test cases that explore diverse combinations of the predicate, such as cases where different subsets of the sub-conditions evaluate to true (e.g., condition 1 and 2, condition 1 and 3, or all conditions simultaneously). As a result, although JaCoCo reports full method-level coverage, some feasible execution paths within the conditional remain unexplored. Importantly, these partially covered branches are not surfaced as actionable targets for further exploration. 

This leads to a key limitation in our approach: no additional path information is exposed in the prompt once a method is deemed fully covered. Consequently, the LLM lacks guidance on which specific paths remain insufficiently tested, preventing further coverage improvement. To address this limitation, future work could incorporate fine-grained coverage feedback, explicitly exposing partially covered branches and their associated path constraints to the prompt. In addition, techniques for systematically enumerating and prioritizing unexplored path combinations could be integrated to guide the LLM toward more diverse and targeted test generation.

\subsection{Unreachable Paths}

\begin{figure}[h!]
    \centering
    \includegraphics[width=1\linewidth]{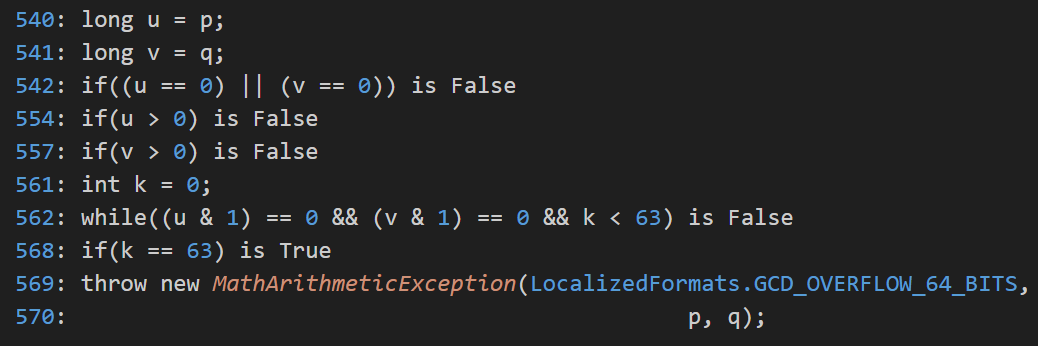}
    \caption{One of the path information for method \texttt{gcd()} in \texttt{ArithmeticUtils.java}}
    \label{fig:ArithmeticUtils_path_example}
\end{figure}

Because JaCoCo reports missed instructions, lines, and branches purely based on runtime execution, infeasible branches or unreachable lines may still be counted as ``missed'' as long as they appear in the compiled bytecode and are included in JaCoCo’s coverage counters. 
In other words, JaCoCo does not distinguish between feasible but uncovered paths and inherently infeasible ones, which can lead to misleading coverage signals.

A concrete example from our study is shown in Figure~\ref{fig:ArithmeticUtils_path_example}, which presents one execution path of the method \texttt{gcd()} in \texttt{ArithmeticUtils.java} from the Math-2f project. Along this path, the variable \texttt{k} is initialized to 0. According to the source code, the only location where \texttt{k} can be updated is within the \texttt{while} loop at line 562. However, because the loop condition at line 562 evaluates to \texttt{false}, the loop body is never executed, and thus \texttt{k} remains unchanged throughout the execution.

As a consequence, the subsequent condition at line 568, whose evaluation depends on \texttt{k} being modified within the loop, becomes unreachable along any feasible execution of this path. Despite this, JaCoCo still marks the corresponding line as partially covered, since one branch of the condition is never observed during execution. This example highlights a fundamental limitation of coverage-based metrics: they may flag certain lines or branches as insufficiently covered even when the missing paths are semantically infeasible, thereby overstating the gap in test adequacy.

\subsection{LLM Capability Limitation}

Although the issue mentioned above may occur in some cases, the main cause of coverage decrease is the model’s limited ability to understand and generate code, which directly affects the quality of the generated test cases. This limitation affects not only the test generator, but also the LLM-based path solver (for object-related paths). 
To better understand these failures, we manually inspected paths that were incorrectly solved by the LLMs, together with some generated test cases, and summarized the major sources of model errors into four categories: 

\begin{itemize}
\item{\textbf{Lack of Code Knowledge:}} The ability to understand code semantics is essential for LLMs to generate effective test cases. We observed several cases where the LLM-based path solver produced incorrect path solutions because it misunderstood the behavior or type relationship of the target code. For example, in \texttt{BasicTypeConverter.java} (JxPath-22f), one path of \texttt{unmodifiableCollection()} is shown in Figure~\ref{fig:unmodifiableCollection_path_example}. For this path, the LLM-based path solver generated \texttt{new HashSet()} as the input. However, since \texttt{HashSet} is an implementation of \texttt{Set}, the condition at line 462 cannot be satisfied. This example shows that the LLM may lack sufficient knowledge of object type hierarchies, which leads to incorrect reasoning about path feasibility.
\begin{figure}
    \centering
    \includegraphics[width=1\linewidth]{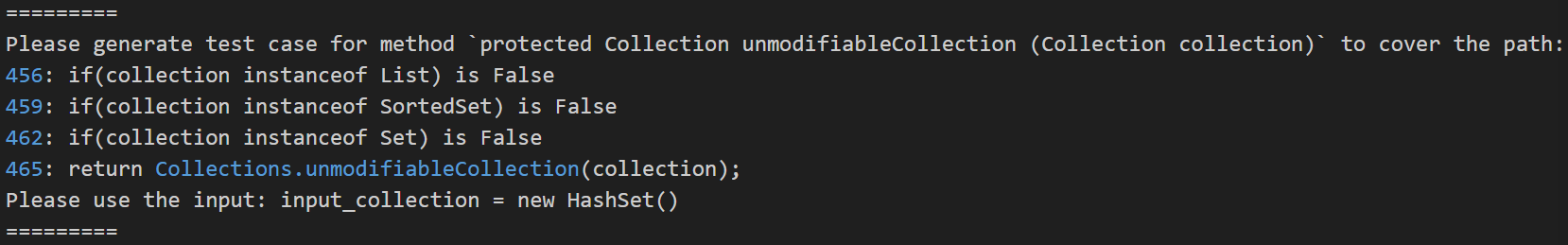}
    \caption{One of the path information for method \texttt{unmodifiableCollection()} in \texttt{BasicTypeConverter.java}}
    \label{fig:unmodifiableCollection_path_example}
\end{figure}

Another example appears in \texttt{BasicTypeConverter.java}, as shown in Figure~\ref{fig:convert_path_example}. In this method, the conditional statement at line 267 requires the return value of another function to be \texttt{null}. Since satisfying this condition requires understanding the behavior of an invoked method beyond the local path itself, the LLM failed to generate an input that reaches the target branch. Although the generated input satisfies all preceding conditions, it does not satisfy the condition at line 267. At the same time, the LLM also hallucinates this as correct. We found that this type of error occurs more frequently when using weaker models (such as \texttt{gpt-4o-mini}).

\begin{figure}[h!]
    \centering
    \includegraphics[width=1\linewidth]{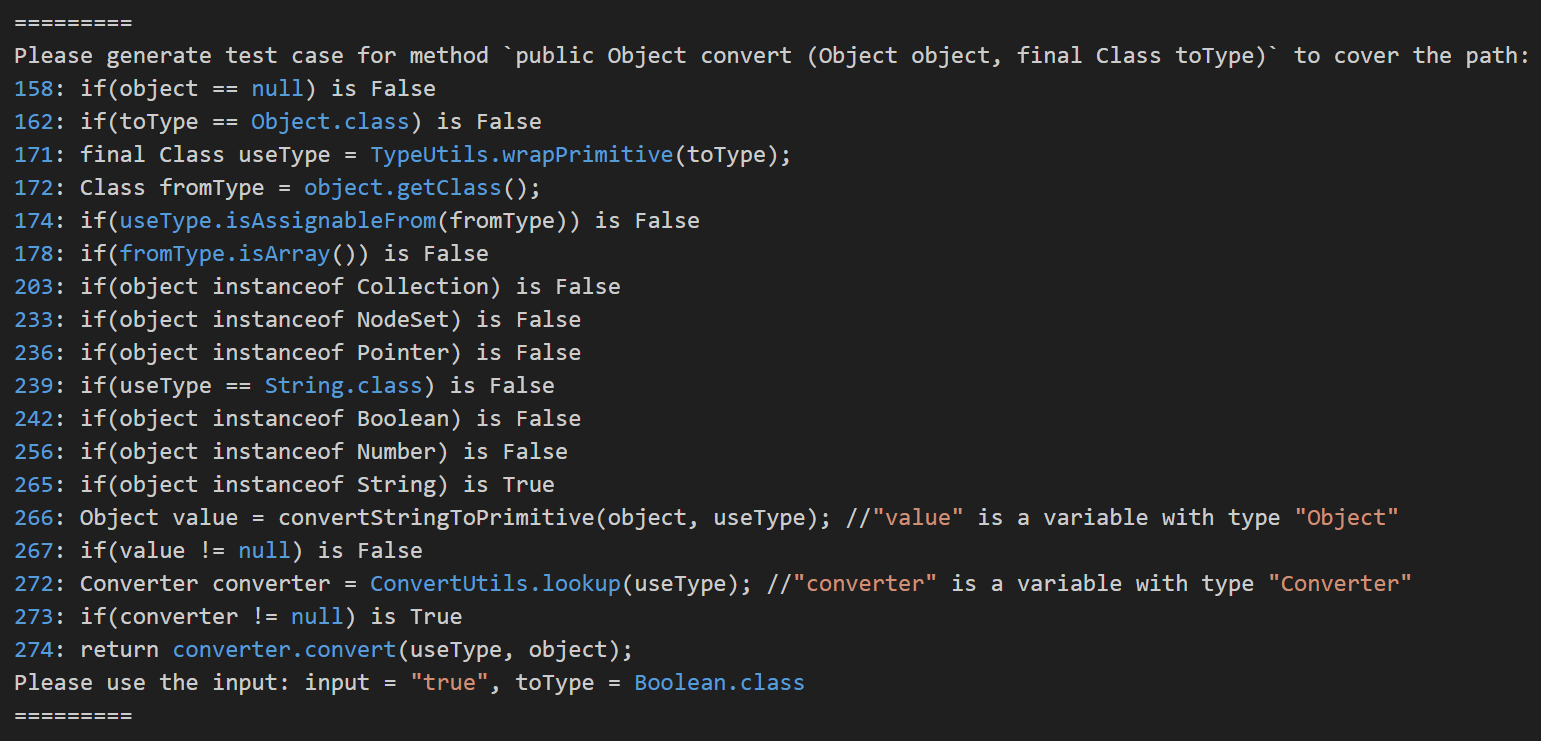}
    \caption{One of the path information for method \texttt{convert()} in \texttt{BasicTypeConverter.java}}
    \label{fig:convert_path_example}
\end{figure}

\item{\textbf{Customized Object Input:}} Our approach only provides the source file that contains the current method under test. However, some methods require input parameters with customized object types defined in other files, which limits the contextual information available to the LLM. For example, in Figure~\ref{fig:decode_path_example}, the \texttt{decode()} method in \texttt{Base32.java} from Codec-18f takes an input parameter of the customized type \texttt{Context}. The definition of this class is located in another Java file and is not included in the prompt. In this case, the LLM has no explicit information about the object structure and can only infer the required input based on the path conditions. In this example, the first condition requires \texttt{context.eof} to be \texttt{false}. However, due to the missing object definition, the LLM hallucinates an input such as \texttt{context.eof = true}. This is not only inconsistent with the required path condition, but also invalid as a test input. As a result, the test generator cannot directly use the provided guidance and may generate a test case based on only partial or incorrect input information.

\begin{figure}[h!]
    \centering
    \includegraphics[width=1\linewidth]{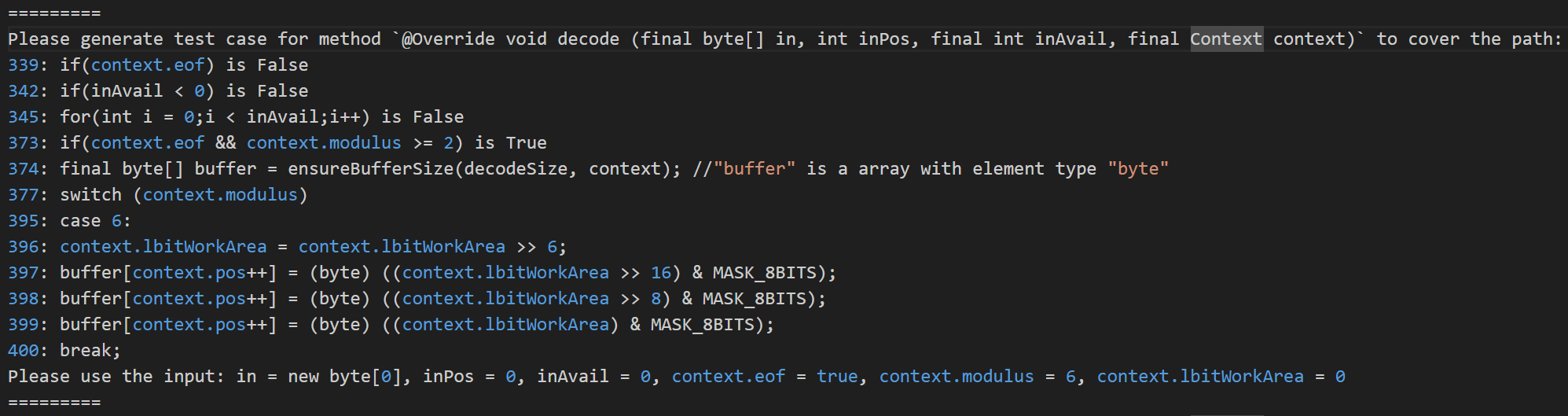}
    \caption{One of the path information for method \texttt{decode()} in \texttt{Base32.java}}
    \label{fig:decode_path_example}
\end{figure}

This situation is more common in complex programs where method inputs depend on project-specific object structures. Therefore, incorporating cross-file context, such as customized class definitions and relevant field information, is an important direction for future improvement.

\item{\textbf{Complex Logic and Code Structure:}} In addition to missing code knowledge and incomplete object context, complex control-flow and code structure can also affect the LLM's ability to correctly reason about the target path. In some cases, the selected path contains multiple method calls, loop statements, object invocations, and several nested or sequential conditional statements. These elements increase the reasoning difficulty because the model needs to simultaneously track variable states, branch outcomes, object behaviors, and the effects of invoked methods across the execution path. When the code is written in a compact (not straightforward) style (such as chained calls, deeply nested conditions, or conditions whose values depend on previous updates), the LLM may overlook some required constraints or incorrectly infer the relationship between statements. As a result, the generated input may satisfy only part of the selected path while failing to meet the requirements of later conditions. This issue shows that, even when the necessary source code is provided, complex logic and code structure can still limit the effectiveness of LLM-based path solving and test generation. And the severity really depends on the ability of the model (e.g., Claude-Sonnet-4.6 will perform better than Claude-3.5-Haiku in code reasoning).

\item{\textbf{Test Case Generation Problem:}} Since we only add test cases that can successfully compile and pass to the test file, the LLM's ability to generate valid and executable tests has a direct impact on the final coverage. In our investigation, we also observed cases where the LLM or symbolic execution produced correct inputs that could cover the target path, but the test generator still produced a failing test case. Even after the test-fixing process, the test could not be repaired and therefore was not added to the test file. According to our path selection mechanism, the same path may not be selected again in the coming iteration. Moreover, if later iterations fail to improve coverage while this path is still not selected again, or if it is selected but the generated tests for this path continue to fail, the iteration limit may be reached, and the program will terminate. In this case, the path remains uncovered even though a correct input solution has already been obtained.

For instance, Figure~\ref{fig:canConvert_path_example} shows one path from the method \texttt{canConvert()} in \texttt{BasicTypeConverter.java} from \texttt{JxPath-22f}. For this path, the input parameter was correctly solved by the LLM-based path solver. 
However, as shown in Figure~\ref{fig:canCovert_test_example}, the generated test still failed after three fixing rounds. The failure was caused by an incorrect assertion, where the correct assertion should be \texttt{assertTrue}. Although this path was selected again in a later iteration, it was still not covered in the final result.

\begin{figure}[h!]
    \centering

    \begin{subfigure}{1\linewidth}
        \centering
        \includegraphics[width=\linewidth]{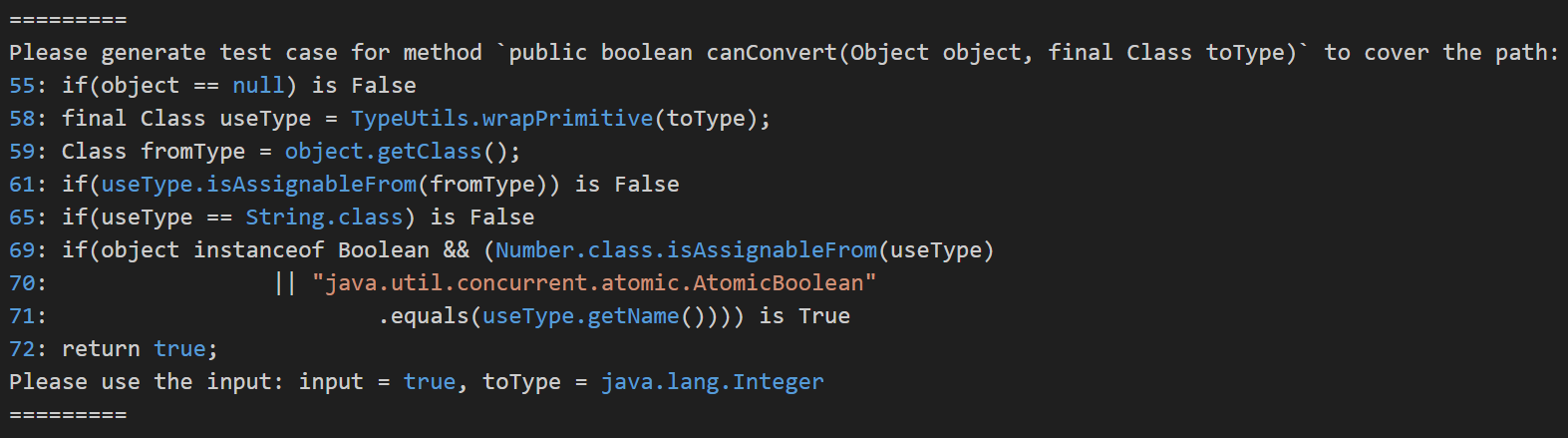}
        \caption{One of the path information for method \texttt{canConvert()}}
        \label{fig:canConvert_path_example}
    \end{subfigure}
    \hfill
    \begin{subfigure}{0.8\linewidth}
        \centering
        \includegraphics[width=\linewidth]{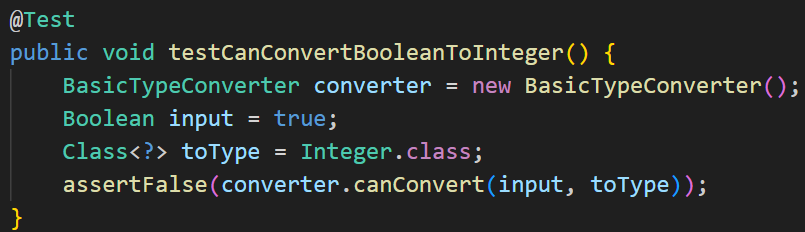}
        \caption{Corresponding test case generated for method \texttt{canConvert()}}
        \label{fig:canCovert_test_example}
    \end{subfigure}

    \caption{One of the path information and the test case generated for method \texttt{canConvert()} in \texttt{BasicTypeConverter.java}}
    \label{fig:canConvert_path_and_test}
\end{figure}

Therefore, correct path solving alone is not sufficient to guarantee coverage improvement. The generated test must also correctly construct the required inputs, invoke the target method, and assert the expected behavior. This indicates that test generation ability remains a critical bottleneck in the overall workflow when using weak LLMs.

\end{itemize}

\section{Impact Statements}

This work demonstrates the value of neuro-symbolic approaches for software test generation by bridging the long-standing gap between precise constraint reasoning and high-quality test synthesis. By combining symbolic execution with large language models (LLMs), {\testgen} shows that it is possible to achieve targeted structural coverage while generating readable, developer-friendly tests, addressing key limitations of both standalone symbolic and LLM-based techniques. Our findings suggest that future testing tools should move beyond purely coverage-driven or generation-driven paradigms and instead incorporate fine-grained, constraint-aware guidance to effectively reach specific program behaviors. Moreover, the results highlight important trade-offs between scalability and precision, indicating that hybrid designs can offer a practical balance for real-world applications. Empirical improvements on Defects4J further demonstrate the effectiveness of this direction, motivating the evolution of benchmarking practices to better capture path-level coverage and usability. Overall, this work provides a foundation for next-generation AI-assisted testing systems that are both semantically grounded and practically deployable, while opening new research directions in constraint-guided generation, feedback-driven refinement, and scalable neuro-symbolic analysis. 
\end{document}